\documentclass[10pt,aps,pra,twocolumn,superscriptaddress,floatfix]{revtex4-1}
\makeatletter
\def\@bibdataout@aps{%
 \immediate\write\@bibdataout{%
  @CONTROL{%
   apsrev41Control,author="08",editor="1",pages="0",title="0",year="1",eprint="1"%
  }%
 }%
 \if@filesw
  \immediate\write\@auxout{\string\citation{apsrev41Control}}%
 \fi
}%
\makeatother 

\usepackage{lipsum}
\usepackage{amsmath}
\usepackage{amsfonts, amssymb, amsthm, bbm, braket, accents}
\usepackage{graphicx}   
\usepackage[usenames,dvipsnames]{color}
\usepackage{soul} 

\newcommand{\ain}{a_{{\rm in}}}
\newcommand{\aout}{a_{{\rm out}}}

\newcommand{\dg}{^\dagger}

\newcommand{\conj}[1]{{#1}^{*}}

\newcommand{\Id}{\openone}

\newcommand{\op}[2]{\left |{#1}\right\rangle\! \!\left \langle {#2}\right |}

\usepackage[breaklinks=true]{hyperref}
\hypersetup{
  colorlinks   = true, 
  urlcolor     = blue, 
  linkcolor    = blue, 
  citecolor   = red 
}
\usepackage{cleveref} 
\crefformat{equation}{Eq.~(#2#1#3)} 
\Crefformat{equation}{Equation~(#2#1#3)}
\crefrangeformat{equation}{Eqs.~(#3#1#4) to~(#5#2#6)}
\crefformat{figure}{Fig.~#2#1#3}

\begin{document}

\title{A passive CPHASE gate via cross-Kerr nonlinearities}

\author{Daniel J. Brod}
\email{dbrod@perimeterinstitute.ca}
\affiliation{Perimeter Institute for Theoretical Physics, 31 Caroline St. N, Waterloo, Ontario, Canada N2L 2Y5}

\author{Joshua Combes}
\email{jcombes@perimeterinstitute.ca}
\affiliation{Institute for Quantum Computing and Department of Applied Mathematics, University of Waterloo, Waterloo, ON, Canada}
\affiliation{Perimeter Institute for Theoretical Physics, 31 Caroline St. N, Waterloo, Ontario, Canada N2L 2Y5}

\date{\today}


\begin{abstract}
A fundamental and open question is whether cross-Kerr nonlinearities can be 
used to construct a controlled-phase (CPHASE) gate. Here we propose a gate
constructed from a discrete set of atom-mediated cross-Kerr interaction sites
with counter-propagating photons. We show that the average gate fidelity $F$ between a CPHASE
and our proposed gate increases as the number of interaction sites increases
and the spectral width of the photon decreases, e.g.\ with 12 sites we find $F > 99\%$.
\end{abstract}


\maketitle

Photons are attractive in quantum information processing as flying qubits and
as a quantum computing platform. To realize the full benefits of quantum
photonic applications, a nonlinearity or photon-photon interaction is usually
required. However, photons only interact in contrived situations
\cite{KarpNeum50}, thus most interactions between photons are {\em effective},
i.e. mediated by matter. For optical quantum computing, in a dual
rail encoding, a natural entangling gate is the controlled-phase (CPHASE) gate
\cite{KokMunrNemo07,NielChua10}.  Unfortunately, the photon-photon
interactions required for a CPHASE gate are hard to engineer. Thus, much of
the progress in the field of optical quantum computing has focused on the KLM
scheme \cite{KnilLaflMilb01} or measurement-based quantum computing
\cite{Niel04,Meni14,GimeShadBrow15}, which circumvent these issues by use of
nondeterministic measurement-induced nonlinearities.

Cross-Kerr interactions have been suggested as a route to a deterministic
Fredkin gate by Milburn \cite{Milb89} and a CPHASE gate by Chuang and
Yamamoto~\cite{ChuaYama95}. These proposals have received less attention than
linear-optical schemes due to two obstacles. First, bulk cross-Kerr
nonlinearities have historically been very small~\cite{VenkSahaGaet13}.
However, experiments in cavity-QED \cite{TurcHoodLang95}, circuit-QED~\cite{HoiKockPalo13}, 
and ensemble systems~\cite{BeckHossDuan15}, have already
demonstrated large cross phase shifts of order one radian per photon.

Second, single-mode analyses fail to account for multimode effects that
preclude a high-fidelity CPHASE gate, as pointed out by Shapiro \cite{Shap06}
and Gea-Banacloche~\cite{GeaBana10}. In principle, a CPHASE gate could be
implemented  by a frequency-local interaction, i.e., with a Hamiltonian
proportional to $a\dg(\omega)a(\omega)b\dg(\omega)b(\omega)$. However,
physically-realistic cross-Kerr effects are {\em spatially} localized, e.g.,
$a\dg(x)a(x)b\dg(x)b(x)$, since they must be mediated by atoms. This creates a
tension between the spectral width of the quanta and the response time of the
Kerr medium. If two temporally broad (spectrally narrow) photons impinge on
the medium, they are likely to both be absorbed by the atoms, but not at the
same interaction site, so no interaction occurs. When temporally narrow
(spectrally broad) photons impinge on the medium the atoms cannot absorb the
photons before they leave the interaction site, and again no interaction
occurs. Shapiro \cite{Shap06}  arrives at similar conclusions, via a
phenomenological model of the  cross-Kerr interaction, that includes a
fidelity-degrading phase-noise \cite{BoivKartHaus94} term. In an intermediate regime, a more
fundamental problem with spatially-local interactions is that they generate
spectral entanglement~\cite{GeaBana10}, e.g.\ when different frequencies 
gather different
cross-phase shifts, or there is frequency mixing. As a consequence of these arguments, it has become
folklore that the multi-mode nature of photons is a fundamental obstacle for
constructing a CPHASE gate from Kerr nonlinearities, even in absentia of other
imperfections.

Here we provide a counter-example to this claim, by constructing a high-fidelity
CPHASE gate using photons that counter-propagate through $N$  atom-mediated 
cross-Kerr interaction sites. In particular, as $N$ increases and the
spectral width of the photons decreases, our proposal tends to a perfect
CPHASE gate. Furthermore, since we do not rely on any phenomenology, our
results unambiguously show that the multimode nature of the field is not a
fundamental obstacle to quantum computation.

There are other proposals for CPHASE gates based on atom-mediated interactions, see 
Refs.~\cite{DuanKimb04,KoshIshiNaka10,ChudChuaShap13,RalpSollMahm15,HackWeltRemp16}. 
Our proposal was motivated by Ref.~\cite{ChilGossWebb13}, where a CPHASE gate was built by a random walk of
counter-propagating qubit waves.  Counter-propagating photonic wave packets, with interactions mediated by Rydberg atoms or
atomic vapours, were investigated in Refs.\  \cite{MassFlei04,FriePetrFlei05,GorsOtteFlei11,GorsOtteDeml10}. 
Our work improves on previous proposals in two ways. First, our construction
requires no active elements, such as error correction, control pulses, switches, or memories. Second, high fidelities 
($F >99\%$) are obtainable with relatively few interaction sites ($N = 12$).

Our main goal is to construct a gate that entangles two qubits encoded in
dual-rail states (see e.g.\ \cite{KokMunrNemo07}) or, equivalently, enact the
two-mode transformation:
\begin{subequations}
\begin{align} 
&\ket{0}_a\otimes\ket{0}_b \rightarrow \ket{0}_a\otimes\ket{0}_b \label{eq:cphasea}\\
&\ket{0}_a\otimes\ket{1_\xi}_b \rightarrow \ket{0}_a\otimes \ket{1_\xi}_b \label{eq:cphaseb}\\
&\ket{1_\xi}_a\otimes\ket{0}_b \rightarrow \ket{1_\xi}_a\otimes\ket{0}_b \label{eq:cphasec}\\
&\ket{1_\xi}_a\otimes\ket{1_\xi}_b \rightarrow e^{i \phi} \ket{1_\xi}_a\otimes\ket{1_\xi}_b, \label{eq:cphased}
\end{align}
\end{subequations}
where $a$ and $b$ are photonic modes, $\ket{0}$ indicates a multimode vacuum,
$\ket{1_\xi}=\int d\omega\, \xi(\omega)a\dg(\omega)\ket{0}$ is a single photon
in the wave packet $\xi (\omega)$, and
$[a(\omega),a\dg(\omega')]=\delta(\omega-\omega')$. Any nontrivial phase
($0<\phi<2\pi$) in \cref{eq:cphased} enables quantum computation, but we are
interested in the case $\phi = \pi$, which corresponds to the CPHASE gate.

To characterize the action of a medium on multimode light, we use the
S-matrix from scattering theory. The S-matrix is a unitary matrix connecting
asymptotic input and output field states i.e. $\ket{\omega_{\rm
out}}=S\ket{\nu_{\rm in}}$, while capturing the relevant effects of the
medium. The {\em ideal} S-matrices corresponding to \crefrange{eq:cphasea}{eq:cphased}
would be $S_{\rm id,1}(\omega_k;\nu_k) = \delta(\omega_k-\nu_k)$, for single-photon 
states, and $S_{\rm id,2}(\omega_a, \omega_b;\nu_a, \nu_b)=
e^{i\phi} S_{\rm id,1}(\omega_a;\nu_a)S_{\rm id,1}(\omega_b;\nu_b)$ for two-photon states, where
input (output) frequencies are denoted by $\nu_{k}$ ($\omega_{k}$), for $k =
\{a,b\}$.  Typically, however, the {\em actual} S-matrices for matter-mediated interactions are of the form $S_{\rm
act,1}(\omega_k;\nu_k)= e^{i\phi_k}\delta(\omega_k-\nu_k)$ and $S_{\rm
act,2}(\omega_a, \omega_b;\nu_a, \nu_b)= S_{\rm act,1}(\omega_a;\nu_a)S_{\rm
act,1}(\omega_b;\nu_b)+ C \delta(\omega_a  + \omega_b - \nu_a - \nu_b)$, where
the coefficient $C$ depends on all frequencies and the parameters of the
interaction mediators \cite{XuRephFan13,BrodComb16a}. The phase $e^{i\phi_k}$ in $S_{\rm
act,1}$ leads to a deformation of the single-photon wave packets, while the function $ \delta(\omega_a  +\omega_b - \nu_a -
\nu_b)$ in $S_{\rm
act,2}$, which arises from energy conservation~\cite{XuRephFan13}, is usually
identified as the source of spectral entanglement.

One important choice we make is to ignore single-photon deformation, which is enforced by mapping all S-matrices as $S\rightarrow S_{\rm
act,1}\dg(\omega_a;\nu_a)S_{\rm act,1}\dg(\omega_b;\nu_b) S$. 
Most previous proposals do not do this (e.g.~\cite{GeaBana10}), which accounts for part of the discrepancy in the maximum fidelities obtained. 
Single-photon deformation could have two negative effects for our proposal. First, it might disrupt linear-optical steps of the computation. This can be avoided by ensuring all photons are deformed  equally at each computational time step \footnote{This can be done e.g.\ by passing photons through empty cavities or uncoupled atoms.}. Second, our results are obtained for specific input wave packet shapes, so single-photon effects could significantly degrade the fidelity of subsequent gates; later, we show that this is not the case for a few rounds of deformation. It is then possible to use measurement-based quantum computing, where each photon experiences at most two CPHASE gates
\cite{ChilLeunNiel05}, or teleportation-based error correction \footnote{ Specifically, photonic qubits which have experienced $M$ rounds of CPHASE gates can be teleported onto wave packets which have only experienced one round of distortion.}.
Finally, it is also possible to physically undo this deformation if necessary, as proposed e.g.\ in Ref.\ \cite{ChudChuaShap13}.

Ideally, we would like to show that the S-matrix for our proposal approaches
$S_{\rm id,2}$ in some limit. However, it is sufficient for this to hold only
for the particular states that we are considering. Thus, to gauge the quality
of our operation, we use the average gate fidelity \cite{Niel02}
\begin{align} \label{eq:fidelity0}
F(\phi) := \int d\psi \bra{\psi} S_{\rm id}(\phi)\dg S_{\rm act}\op{\psi}{\psi}  S_{\rm act}\dg S_{\rm id}(\phi) \ket{\psi}
\end{align}
where the integration is taken over the Haar measure of the joint Hilbert
space (for further details, see \Cref{apx:SSC}). For our gate to be useful for
quantum computation, it suffices that $F = 1-\epsilon$, where $\epsilon$ is
some constant threshold \cite{SandWallSand16}. 

\begin{figure}[t]
\includegraphics[width=0.85\columnwidth]{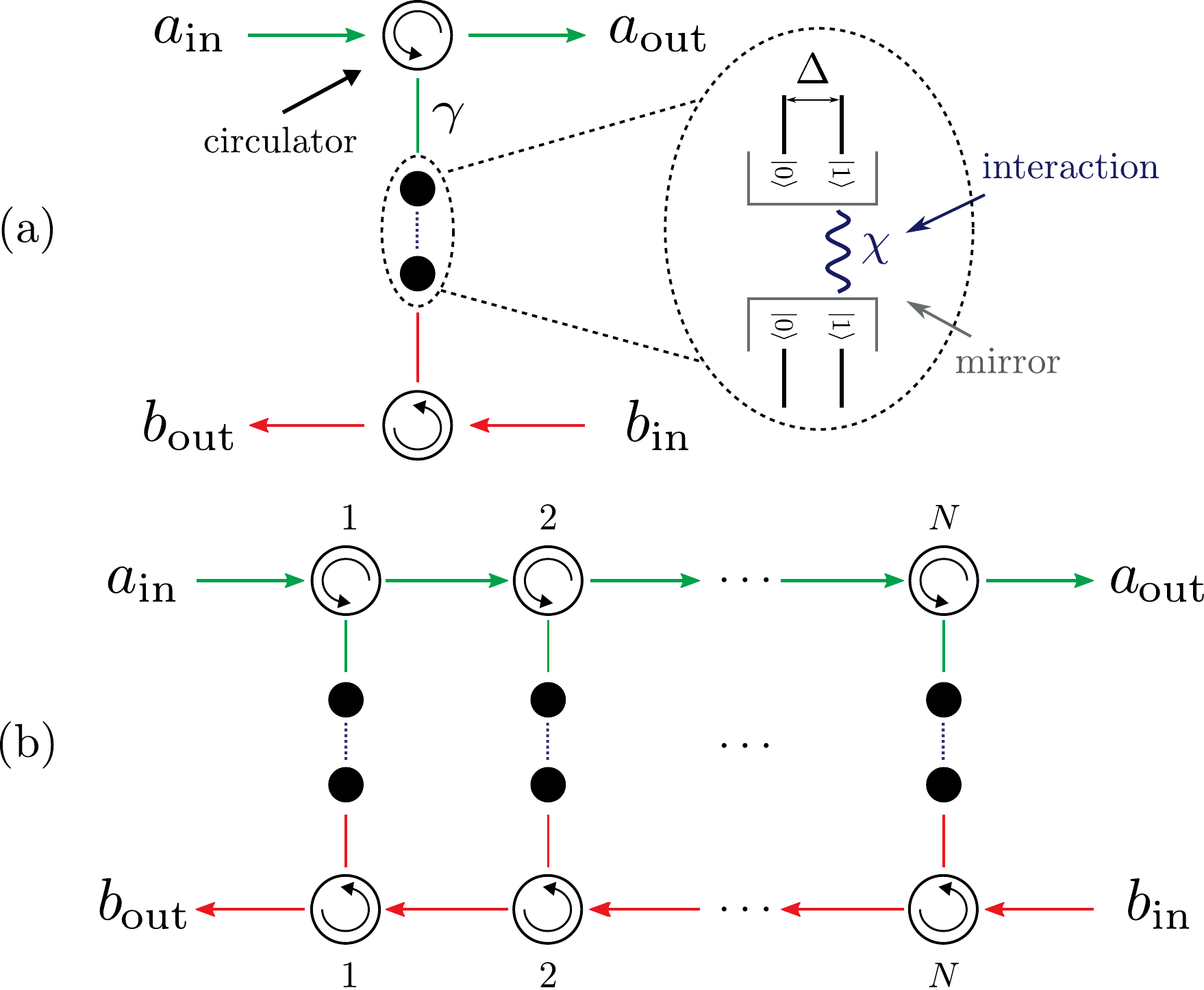}
\caption{(Color Online) 
(a) The physical system inside our unit cell. It consists of two coupled two-level atoms, with internal energies $\Delta$, and which interact via
$H=\chi (\Id- A_{z})(\Id- B_{z})= \chi \op{1,1}{1,1}$. The input-output fields couple to the atoms via the relation
$\aout = \sqrt{\gamma}A_{-} +\ain$, and similarly for mode $b$. It was shown in Ref.\ \cite{BrodComb16a} that this system gives rise to the same S-matrices for single- and two-photon
scattering as a pair of crossed cavities with cross-Kerr interaction between them. In the limit $\chi \rightarrow \infty$, this reduces to 
a three-level atom in a ``V'' configuration, such as considered in Ref.\ \cite{ChudChuaShap13}. (b) Our main proposal using $N$ discrete interaction sites with counter-propagating photons.}\label{fig1}
\end{figure}

\begin{figure*}[ht!]
\includegraphics[width=\textwidth]{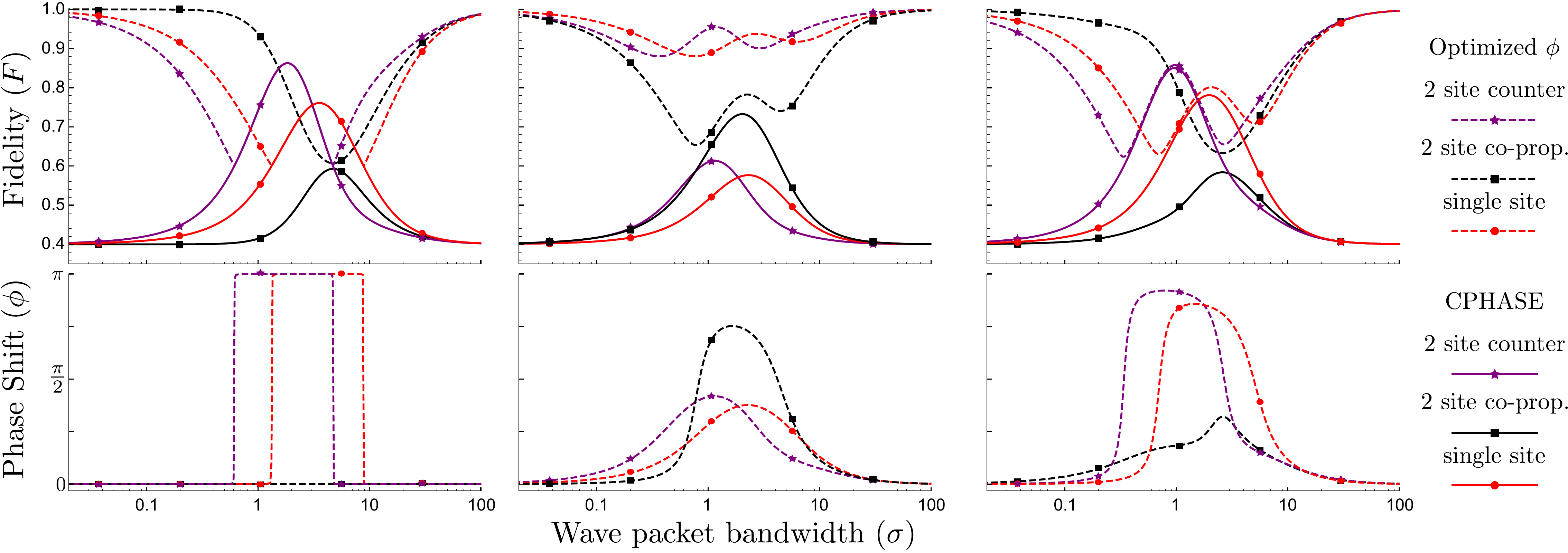}
\caption{(Color online). 
In the top row, solid
lines represent the average gate fidelity with respect to the CPHASE gate, i.e.
$F(\phi=\pi)$, while dashed lines denote the fidelity $F(\phi)$ maximized with respect to some $\phi$. The second row plots the corresponding phase shift $\phi_{\rm
opt} = {\rm argmax}_\phi\,F (\phi)$. 
We compare three cases of interest: (i) a single-site Kerr interaction (circles),
(ii) a two-site interaction with co-propagating photons (squares), (iii) the
two-site interaction with counter-propagation (stars).
In column 1, we have chosen
$(\omega_0,\gamma,\chi) = (0,10,10000)$, which maximizes $F$ for the counter-propagating 
case, resulting in $F_{\rm counter}= 0.8628$ (only for this case, whenever $F \gtrsim 0.6$, the dashed and solid lines coincide). 
In column 2 we chose $(\omega_0,\gamma,\chi) = (0,6,2.67)$ to maximize $F$ for
the co-propagating case obtaining $F_{\rm co-prop.}= 0.7326$, and in column 3 we
chose $(\omega_0,\gamma,\chi) = (1.1,4.5,5)$ to optimize the single-site $F$,
obtaining $F_{\rm single}=0.7810$.} \label{fig2}
\end{figure*}

\textit{Single- and two-site gate fidelities.} We begin by examining $F$ for wave packets scattering from a
single site, as well as two sites in a co- and counter-propagating arrangement.
The discrete Kerr interaction we consider is depicted in \cref{fig1} (a). The unit cell we repeat in \cref{fig1} (b), call it $G=(L,H)$, can be described using  
``LH"~\cite{Carm93,Gard93,GougJame09} parameters from input-output theory, where $L$ is a vector of operators that couple the field to the system and $H$ is the system Hamiltonian. The  LH parameters for our unit cell are
\begin{align}
G
= & \bigg[ \left(\begin{array}{c}\sqrt{\gamma} A_{-} \\ \sqrt{\gamma} B_{-}\end{array}\right), \frac{\Delta}{2} (\Id - A_{z}) + \frac{\Delta}{2} (\Id - B_{z}) \notag \\ & + \chi (\Id- A_{z}) (\Id- B_{z}) \bigg],\label{unitCellSLHModel}
\end{align}
where $A_{-}$ and $ A_{z}$ are the atomic lowering and Pauli Z operators for
atom $A$, and likewise for $B_{-}$ and $B_{z}$.  
We cascaded $N$ unit cells with co- and counter-propagating fields and computed the corresponding 
S-matrices, as detailed in Ref.~\cite{BrodComb16a} and which we reproduce in \Cref{apx:Smatrices}. The final ingredient needed to
calculate $F$ is the wave packet shape, which we choose to be Gaussian with
detuning $\omega_0$ (i.e.\ carrier frequency $\omega_c=\Delta+\omega_0$) and bandwidth $\sigma$, i.e.\
\begin{align}\label{eq:gaussianphoton}
\xi(\omega) = \frac{1}{(2\pi \sigma^2)^{1/4}}\exp\left[ -\frac{(\omega -\omega_c)^2}{4 \sigma^2}  \right].
\end{align}

In \cref{fig2}, column 1, we display the plots for parameters $(
\omega_0,\gamma,\chi)$ which maximize the fidelity of counter-propagating wave
packets. 
Relative to the single site, we observe a clear increase of the
maximum obtainable fidelity when the photons are counter-propagating, and a
{\em decrease} when they are co-propagating, as illustrated in the top row. In the limit of large
$\chi$ and $\omega_0=0$, the phase shift is always either $0$ or $\pi$,
corresponding to the identity or CPHASE gate respectively (see the second row).
We observe that counter-propagating wave packets
tend to perform better than co-propagating for a large region of the parameter
space, but there are exceptions.

In \cref{fig2}, column 2, we display a parameter regime where co-propagating
photons obtain their maximum fidelity and outperform the other two cases. The
explanation for this is the following. In this regime, the co-propagating case
seems to suffer more spectral entanglement, but also acquire a larger phase
shift, than the other two (see the dashed lines), and the tradeoff between these effects leads to a
higher fidelity with the CPHASE gate. However, these effects are linked in
such a way that this peak fidelity and the maximum phase are still much
inferior to the best obtained by the single-site and counter-propagating cases
in other parameter regimes. Nonetheless, this suggests it is possible to
use a perturbative approach to construct long weakly-coupled atom
chains where the rate at which phases and spectral entanglement accumulate are
more benign (e.g.\ \cite{ChudChuaShap13}). It is also interesting that the
peak of the fidelity in the co-propagating case happens for larger $\sigma$
than for the counter-propagating case in all three columns, which could lead to
a CPHASE gate for spectrally broader photons.

In \cref{fig2}, column 3, we display parameters that maximize the single-site
fidelity. As we generically expect, the counter-propagating wave
packets outperform the single-site and co-propagating ones both in fidelities
and phase shifts. This happens even when $\omega_0 \neq 0$,
indicating that our conclusions are somewhat robust with respect to being off-resonance.

\textit{$N$-site gate fidelities.} We now investigate the average fidelity of our proposal to the CPHASE gate as
we increase the number of interaction sites. Based on observations from the two-site case,
we restrict our analysis to
counter-propagating photons, working on-resonance ($\omega_0=0$), and take a
$\chi\rightarrow \infty$ limit, since this yields the most promising results. We
also take $\gamma=1$, since choosing other values 
effectively rescales $\sigma$ when working on-resonance. Thus, the average
gate fidelity $F$ is a function of the number of interaction sites $N$ and the
photon bandwidth $\sigma$: $F(\sigma,N)$.

\begin{figure}[th!]
\includegraphics[width=\columnwidth]{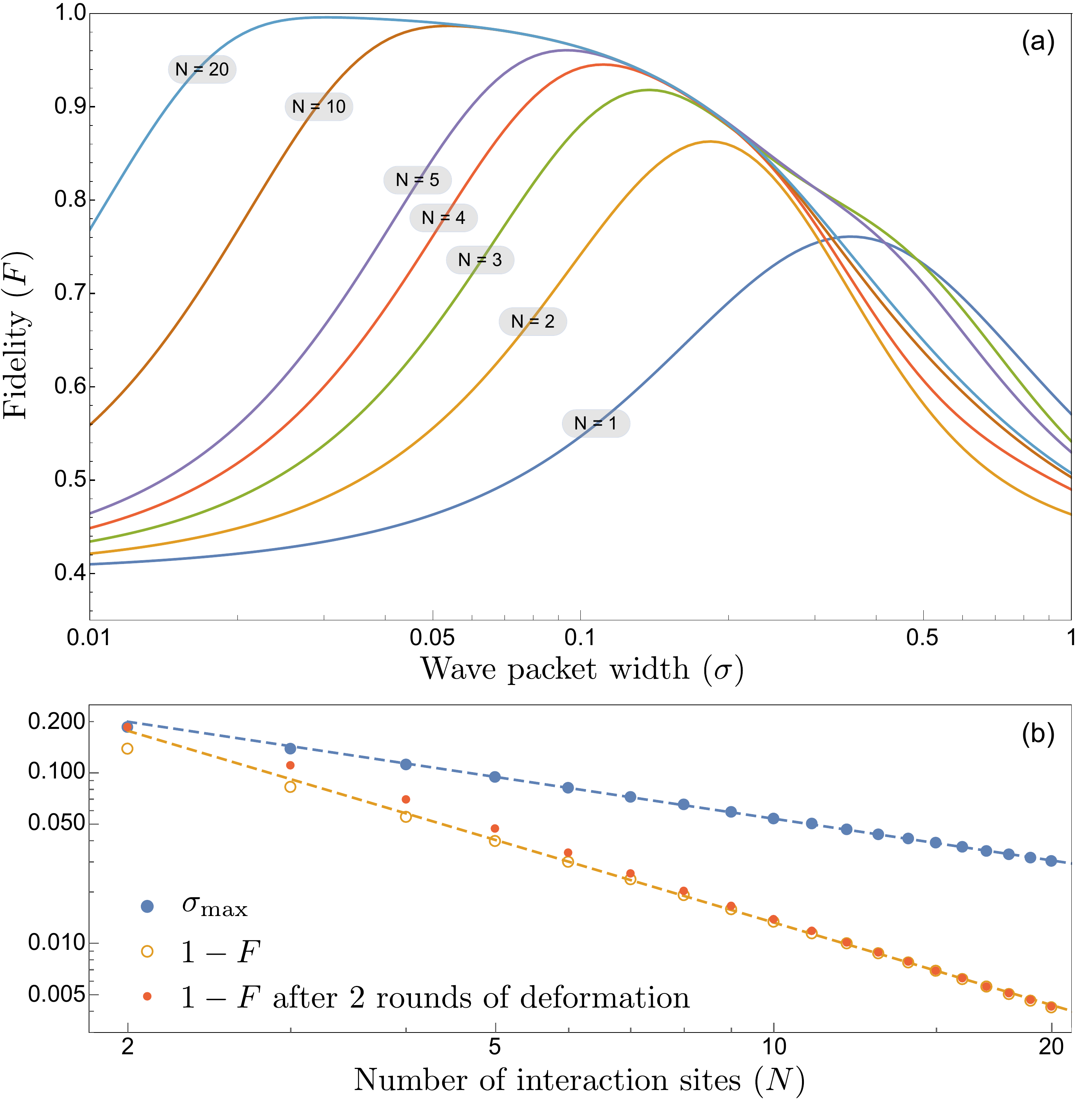}
\caption{(Color online). (a) Average gate fidelity between our proposal and 
the CPHASE as a function of frequency bandwidth $\sigma$ for increasing number 
of interaction sites $N$. (b) We take the maximum of the gate fidelity in (a), 
and plot the infidelity $(1-F)$ and the corresponding maximizing 
$\sigma_{\rm max}$ as functions of the number of interaction sites $N$. Small
red dots correspond to $1-F$ for photons that have undergone two rounds of single-photon
deformation. The dashed lines correspond to the fits $1-F(\sigma_{\rm max},N)=0.537 N^{-1.61}$ 
and $\sigma_{\rm max}=0.350 N^{-0.81}$, where we fit to $N\in [4,20]$.} \label{fig3}
\end{figure}

In \cref{fig3}(a) we plot the average gate fidelity, as a function of
$\sigma$, for increasing $N$. Notice that, as the number of interaction sites
increases, the  maximum average fidelity increases, indicating that the
resulting operation is sequentially closer to a CPHASE gate. Also notice that,
besides attaining higher maximum values, the fidelity curve is also  becoming
broader (albeit in logarithmic scale). This means that, as the number of sites
is increased, the proposed CPHASE gate becomes more broadband, or robust with
respect to the spectral bandwidth of the photon. The highest value for the fidelity
in \cref{fig3}(a) is $0.996$, when $N=20$.

In \cref{fig3}(b), we investigate the maximum of the average fidelity $F_{\rm
max}$ and its corresponding value of $\sigma_{\rm max}$ as a function of $N$.
We see that $1-F_{\rm max}$ is monotonically decreasing and $\sigma_{\rm max}$
slowly tends towards the plane-wave limit. For the observed behaviour of
\cref{fig3}(a-b) we predict that, in order to obtain $F_{\rm max} =0.999$, we
would need $N\simeq 50$ and $\sigma\simeq 0.014 s^{-1}$. \Cref{fig3}(b) also
shows that, for $N > 5$, the fidelity is not significantly affected by using single-photon wave packets that have
suffered one or two rounds of deformation.

\begin{figure}[t!]
\includegraphics[width=\columnwidth]{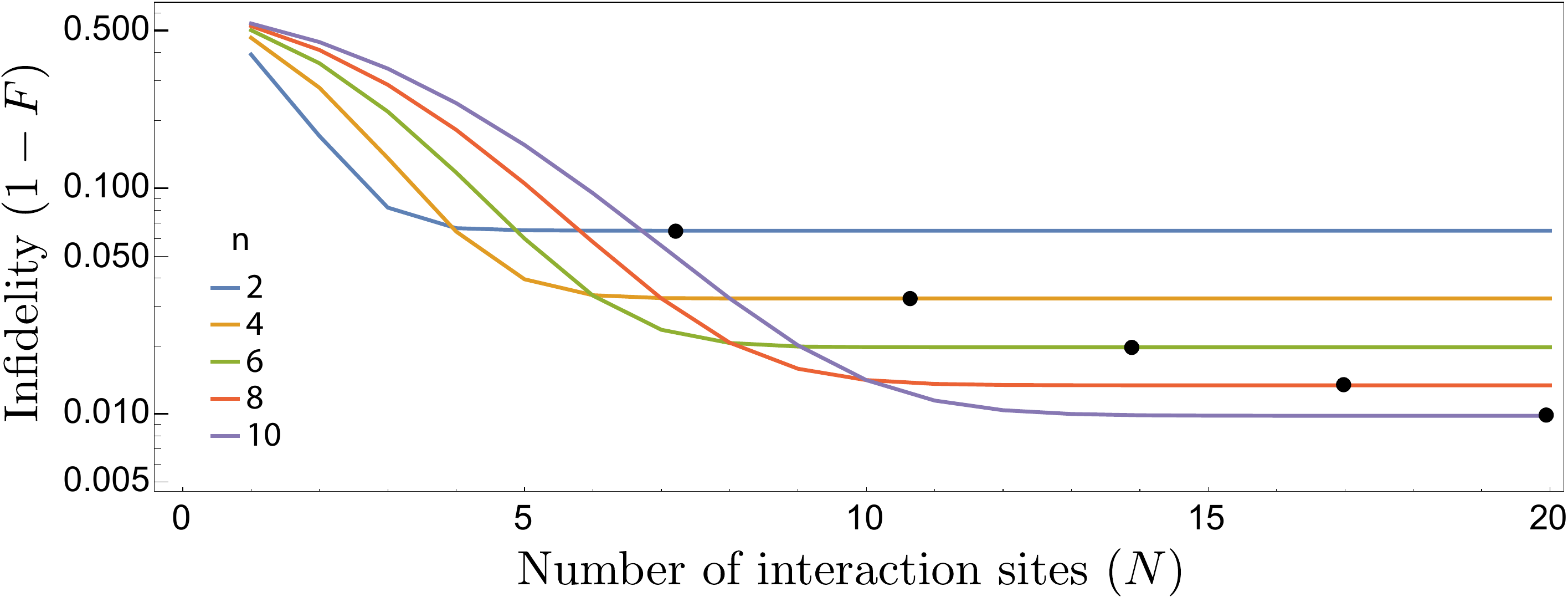}
\caption{(Color online). Saturation of the average gate infidelity for fixed 
$\sigma$ as the number of interaction sites $N$ increases. For illustration, we 
choose the $\sigma$'s that maximize the fidelity for specific numbers of 
interaction sites $n$, i.e. $\sigma_{\rm max}^n={\rm argmax}_\sigma\, F(\sigma,n)$ where $n = \{2, 4, 6, 8, 10\}$. The lines then correspond to the fidelities at 
$\sigma_{\rm max}^n$ for increasing $N$, i.e.\ $F(\sigma_{\rm max}^n,N)$. Finally, 
the black dots in each curve correspond to $N=1/\sigma_{\rm max}^n$, where we 
predict the fidelity to saturate.}\label{fig4}
\end{figure}

Another feature apparent in \cref{fig3}(a) is that, for fixed $\sigma$, the
advantage gained from increasing $N$ eventually saturates. This is explored
further in \cref{fig4}. An intuitive explaination is: interpret $1/\gamma$ as the
typical timescale before an excited atom re-emits a photon, then $t_m \approx
N/\gamma$ is the time that each wave packet remains inside the medium. Thus, 
if the wave packets have temporal width of $t_w\approx 1/\sigma$,
when $N$ is roughly $\gamma/\sigma$ the chain becomes ``long enough'' to
contain the entire wave packets, and the interaction saturates.

Our results show that, to obtain higher-fidelity gates, one has to move to smaller values of
$\sigma$ together with longer atomic chains. In fact, in Ref.~\cite{BrodComb16a}, 
we and Gea-Banacloche have shown that, in the limit where $\sigma \rightarrow 0$ and $N \rightarrow \infty$,
the S-matrix for the $N$-site case tends to the ideal one (modulo single-photon deformation) $S_{\rm 2}(\omega_a, \omega_b;\nu_a, \nu_b)= - S_{\rm act,1}(\omega_a;\nu_a)S_{\rm act,1} (\omega_b;\nu_b)$.
 This is independent of the specific wave packet shape, further motivating our choice to ignore single-photon deformation. The results presented here
are more relevant for implementations, as one only needs to
increase $N$ and decrease $\sigma$ until the fidelity surpasses the
threshold necessary for fault-tolerant computation.

\textit{Discussion.} Our goal was to determine if it possible to build a passive CPHASE gate using
cross-Kerr interactions, and we have shown that it is. Importantly, our
results do not contradict those of Ref.~\cite{Shap06}, which uses a
phenomenological model of a cross-Kerr medium. Using that model,
a CPHASE gate might indeed be unachievable. However, our results are
based on a fully multi-mode treatment of the field and a fully microscopic
treatment of the interaction mediators. Thus we believe they do provide a
counter-example against the stronger claim, which is frequently propagated the literature, that the multimode nature of the field is a
fundamental physical obstacle to implementing a CPHASE gate. Furthermore, our
proposal enjoys two advantages over prior proposals. First, our gate is
passive, i.e., it does not require active error correction, such as the
principal mode projection technique used in Ref.~\cite{ChudChuaShap13}.  Also, 
our proposal requires fewer interaction sites to achieve a fixed fidelity, e.g. 
in Ref.~\cite{ChudChuaShap13} the authors estimate they  
need $10^6$ interaction sites for a $95\%$ fidelity with a CPHASE gate, whereas our proposal
achieves that value with 5 sites.

Admittedly, our proposal is a proof-of-principle result that will be
challenging to construct in practice. Further, we hope our construction will
inspire others to devise simpler and less resource-intensive proposals. There
are plenty of avenues to explore, e.g.\ placing the atomic interaction sites
inside cavities to gain a cavity enhancement \cite{FanJohaComb14}, or varying the atom parameters
along the chain~\cite{HushCarvHedg13} while simultaneously varying the input photon
wave packet shape. Our analysis did not include any additional imperfections,
and we leave as future work to adapt our model to include other effects such
as losses, emission into non-guided modes, coupling to a thermal bath, etc. 
In \Cref{apx:problem_and_rules} we propose an update to a set of rules, initially
laid out by Gea-Banacloche \cite{GeaBana10}, that must be satisfied by any theoretical proposal 
for a realistic CPHASE gate,
based on conclusions drawn from this work and \cite{BrodComb16a}.

 The authors acknowledge helpful discussions with Agata Bra\'nczyk, Daniel
 Gottesman, Bing He, Raissa Mendes, Barry Sanders, and Zak Webb. The authors also thank Julio 
 Gea-Banacloche and Jeffrey Shapiro for feedback and friendly discussions about
 this work. 
{This research was supported by Perimeter Institute for Theoretical Physics. Research at Perimeter Institute is supported by the Government of Canada through the Department of Innovation, Science and Economic Development Canada and by the Province of Ontario through the Ministry of Research, Innovation and Science.}


\bibliography{../../scattering_ref}


\newpage
\onecolumngrid

\newpage

\appendix
\section{Average gate fidelity}\label{apx:SSC}

In this paper, we gauge the quality of our operation using the standard
average gate fidelity \cite{Niel02}:
\begin{align} \label{eq:fidelity0b}
F_1(\phi) := \int d\psi \bra{\psi} S_{\rm id}(\phi)\dg S_{\rm act}\op{\psi}{\psi}  S_{\rm act}\dg S_{\rm id}(\phi) \ket{\psi}
\end{align}
where the integration is over the two-qubit Haar measure, and $S_{\rm id}$ and
$S_{\rm act}$ include the removal of the single-photon deformation. \Cref{eq:fidelity0b} is the same as \cref{eq:fidelity0} in the main text.
In order
to carry out the averaging explicitly, one parameterizes $\ket{\psi}$ as
\cite{KendNemoMunr02}
\begin{align*}
\ket{\psi} = e^{i \chi_0} \cos{\theta_0} \ket{00} + e^{i \chi_1} \sin{\theta_0} \cos{\theta_1} \ket{01} + e^{i \chi_2} \sin{\theta_0} \sin{\theta_1} \cos{\theta_2} \ket{10} +e^{i \chi_3} \sin{\theta_0} \sin{\theta_1} \sin{\theta_2} \ket{11},
\end{align*}
where $0 \leq \chi_i < 2\pi$ and $0 \leq \theta_i < \pi/2$. In this
parameterization, the Haar measure in the integration is
\begin{equation*}
d\psi = 48(2\pi)^{-4} (\sin{\theta_0})^5 \cos{\theta_0} (\sin{\theta_1})^3 \cos{\theta_1} (\sin{\theta_2})^5 \cos{\theta_2} d \theta_0 d\theta_1 d\theta_2 d \chi_0 d \chi_1 d \chi_2 d \chi_3.
\end{equation*}
Since our analysis removes all single-photon deformation, we can leverage the
fact that the only state affected by the interaction is
$\ket{1_\xi}_a\otimes\ket{1_\xi}_b$ and that our gate conserves photon number
to write
\begin{equation} \label{eq:fidelity3}
F_1(\phi) = \frac{1}{10} \left( 6 + 3 \textrm{Re}(e^{i \phi} \mathcal{F}) + |\mathcal{F}|^2 \right).
\end{equation}
Here $\mathcal{F}$ is the overlap between the single- and two-photon wave packets:
\begin{equation}
\mathcal{F} = \int \conj{[\xi^{\rm out}_1(\nu_a,\nu_b)]} \xi^{\rm out}_2(\nu_a,\nu_b) d\nu_a d\nu_b,
\end{equation}
where 
\begin{equation}
\xi_i^{\rm out}(\nu_a,\nu_b) = \int {\xi^{\rm in}(\omega_a) \xi^{\rm in}(\omega_b) \bra{\nu_a \nu_b} S_{i} \ket{\omega_a \omega_b} } d\omega_a d\omega_b
\end{equation}
for $i=1,2$ are the propagated two-photon wave packets, respectively,
according to the S-matrices computed in Ref.~\cite{BrodComb16a}.

Often, $\mathcal{F}$ is used as the main figure of merit, since it relates the
transformed two-photon wave packet to two copies of a single-photon wave
packet, thus directly measuring undesired effects such as spectral
entanglement. However, we believe that the average fidelity, for our purposes,
is a more transparent and unambiguous figure of merit, because it determines
how well one approximates  a desired gate in the computational state space.
Although the average gate fidelity is not the figure of merit that appears in
the threshold theorem directly, these quantities can be related \cite{SandWallSand16}.

\begin{figure}[h]
\includegraphics[width=0.6\columnwidth]{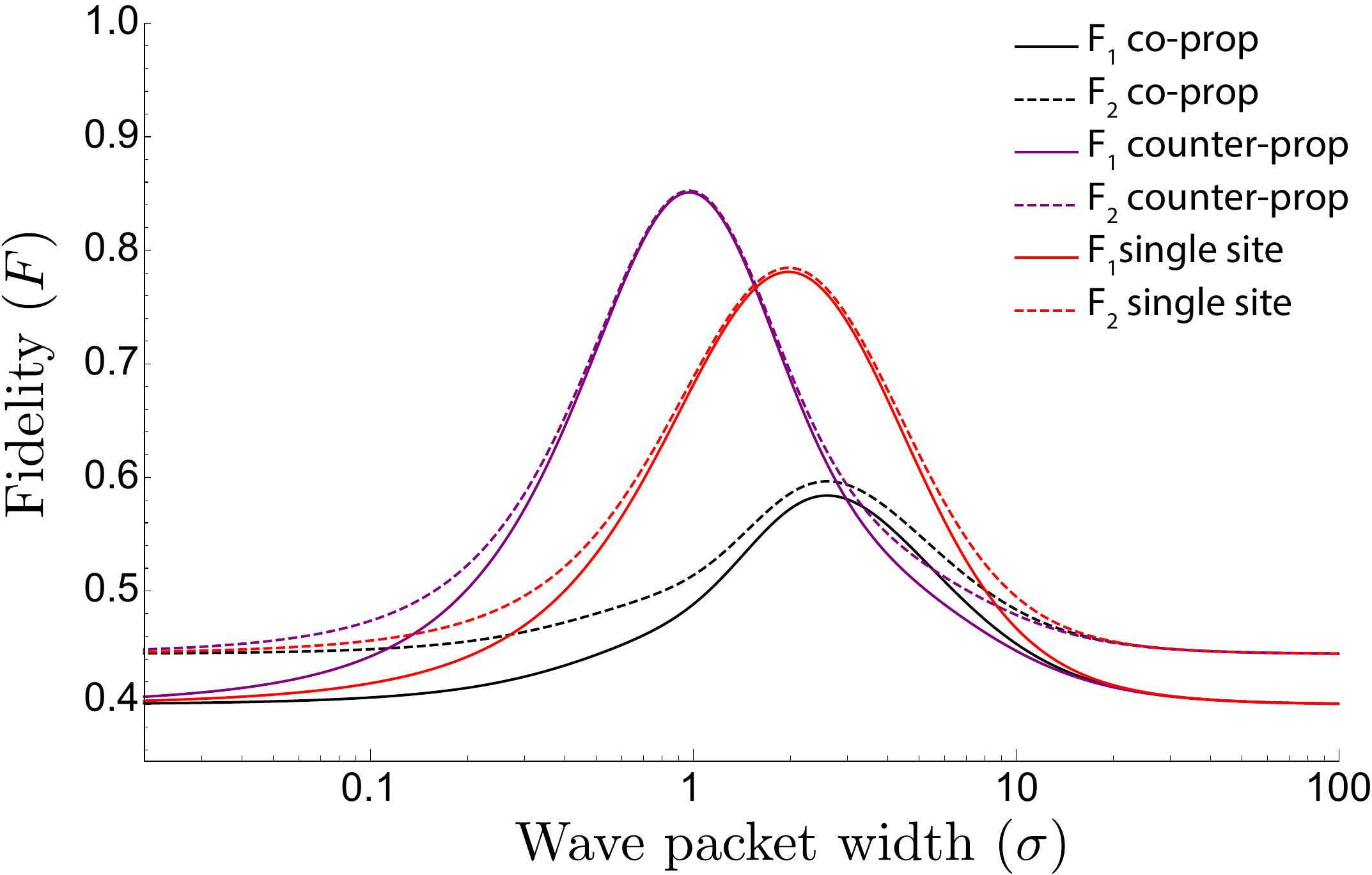}
\caption{(Color online). Comparison between average fidelities using a Haar average 
over full two-qubit Hilbert space ($F_1$) and over the two single-qubit Hilbert spaces ($F_2$).
Here $\chi= 5$, $\gamma= 4.5$, and $\omega_0 =1.1$.} \label{fig5}
\end{figure}

We should also point out that Shapiro, in Ref.~\cite{Shap06}, considers a slightly
different figure of merit, where the average in \cref{eq:fidelity0b} is done
over all {\em product} two-qubit states. Using that definition, the state
$\ket{\psi}$ can be parameterized as
\begin{align*}
\ket{\psi} = (\cos{\theta_0} \ket{0} + e^{i \chi_0} \sin{\theta_0} \ket{1})(\cos{\theta_1} \ket{0} +e^{i \chi_1}\sin{\theta_1} \ket{1}),
\end{align*}
with the Haar measure given by
\begin{equation*}
d\psi = 4 (2\pi)^{-2} \sin{\theta_0} \cos{\theta_0} \sin{\theta_1} \cos{\theta_1} d \theta_0 d\theta_1 d \chi_0 d \chi_1,
\end{equation*}
and our \cref{eq:fidelity3} would become
\begin{equation} \label{eq:fidelity4}
F_2(\phi) = \frac{1}{18} \left( 11 +5 \textrm{Re}(e^{i \phi} \mathcal{F}) + 2 |\mathcal{F}|^2 \right).
\end{equation}
We have found that both definitions lead to essentially the same results.
This is illustrated in \Cref{fig5}, where we see that $F_1<F_2$ because $F_2$ is less 
conservative. As $\mathcal{F}$ approaches $-1$ (which would correspond to a
perfect CPHASE gate) the difference between $F_1(\pi)$ and $F_2(\pi)$
becomes negligible.  We use \cref{eq:fidelity3} because of its connections to the threshold 
theorem, and because it can capture other undesired effects such as e.g.\ entanglement breaking.

\section{S-matrices}\label{apx:Smatrices}
For completeness, in this Appendix we write the S-matrices computed in Ref.~\cite{BrodComb16a}.
We begin by defining the shorthands
\begin{align*} 
\Gamma(\omega) := \frac{\gamma}{2} + i (\Delta-\omega). \\
\Gamma_i(\omega) := \frac{\gamma_i}{2} + i (\Delta_i-\omega_i),
\end{align*}
which will be used when we have a single and multiple sites, respectively. Although we computed some of these S-matrices with different parameters $(\gamma_i,\Delta_i,\chi_i)$ for each site, in the numerical work reported in the main text we assume them to be equal at all sites for simplicity.

\subsubsection{Single site}
For a single interaction site, the S-matrices for co- and counter-propagating photons are equivalent. In this case, we have the single-photon S-matrix
\begin{equation*}
S_{\rm act,1}(\omega_k;\nu_k) = - \frac{\conj{\Gamma}(\omega_k )}{\Gamma(\omega_k )} \delta(\omega_k  - \nu_k).
\end{equation*}
for $k=a,b$. The two-photon S-matrix is
\begin{align*}
S_{\rm act,2}(\omega_a, \omega_b;\nu_a, \nu_b)=  S_{\rm act,1}(\omega_a;\nu_a)S_{\rm act,1}(\omega_b;\nu_b) - i \frac{\chi \gamma^2}{\pi} \left( 1+\frac{2 i \chi}{\Gamma(\omega_a )+\Gamma(\omega_b)} \right)^{-1} \frac{\delta(\omega_a  + \omega_b - \nu_a - \nu_b)}{\Gamma(\nu_b)\Gamma(\nu_a) \Gamma(\omega_b) \Gamma(\omega_a )},
\end{align*}

\subsubsection{Two sites, co-propagating photons}

For two interaction sites in the co-propagating arrangement, we have the single-photon S-matrix
\begin{equation*}
S_{\rm act,1}(\omega_k;\nu_k) = \frac{\conj{\Gamma_2}(\omega_k ) \conj{\Gamma_1}(\omega_k )}{\Gamma_2(\omega_k )\Gamma_1(\omega_k )} \delta(\omega_k  - \nu_k)
\end{equation*}
for $k=a,b$. The two-photon S-matrix is
\begin{align*}
S_{\rm act,2}(\omega_a, \omega_b;\nu_a, \nu_b)= & S_{\rm act,1}(\omega_a;\nu_a)S_{\rm act,1}(\omega_b;\nu_b) -  \frac{\delta(\omega_a  + \omega_b - \nu_a - \nu_b)}{\pi} \notag \\ 
& \times \bigg[  i \left( \frac{\conj{\Gamma}_2(\omega_a) \conj{\Gamma}_2(\omega_b)}{\Gamma_2(\omega_a) \Gamma_2(\omega_b)} \right) \left( 1 + \frac{2 i \chi_1}{\Gamma_1(\omega_a) + \Gamma_1(\omega_b)} \right)^{-1} \frac{\chi_1 \gamma_1^2}{\Gamma_1(\nu_b)\Gamma_1(\nu_a) \Gamma_1(\omega_b) \Gamma_1(\omega_a )} \notag \\
& + i \left( \frac{\conj{\Gamma}_1(\omega_a) \conj{\Gamma}_1(\omega_b)}{\Gamma_1(\omega_a) \Gamma_1(\omega_b)} \right) \left( 1 + \frac{2 i \chi_2}{\Gamma_2(\omega_a) + \Gamma_2(\omega_b)} \right)^{-1} \frac{\chi_2 \gamma_2^2}{\Gamma_2(\nu_b)\Gamma_2(\nu_a) \Gamma_2(\omega_b) \Gamma_2(\omega_a )} \notag \\
& + \left( 1 + \frac{2 i \chi_1}{\Gamma_1(\omega_a) + \Gamma_1(\omega_b)} \right)^{-1} \left( 1 + \frac{2 i \chi_2}{\Gamma_2(\omega_a) + \Gamma_2(\omega_b)} \right)^{-1} \frac{4 \chi_1 \chi_2 \gamma_1^2 \gamma_2^2}{\Gamma_1(\nu_b)\Gamma_1(\nu_a) \Gamma_2(\omega_b) \Gamma_2(\omega_a )} \notag \\
& \times \frac{1}{(\Gamma_1(\omega_a) + \Gamma_1(\omega_b))(\Gamma_1(\omega_a) + \Gamma_2(\omega_b))(\Gamma_2(\omega_a) + \Gamma_2(\omega_b))} \bigg].
\end{align*}

\subsubsection{Two sites, counter-propagating photons}

For two interaction sites in the counter-propagating arrangement, the single-photon S-matrix is obviously the same as for the co-propagating arrangement. The two-photon S-matrix, however, is
\begin{align*}
S_{\rm act,2}(\omega_a, \omega_b;\nu_a, \nu_b)= & S_{\rm act,1}(\omega_a;\nu_a)S_{\rm act,1}(\omega_b;\nu_b) -  \frac{\delta(\omega_a  + \omega_b - \nu_a - \nu_b)}{\pi} \notag \\ 
& \times \bigg[  i \left( \frac{\conj{\Gamma}_2(\omega_a) \conj{\Gamma}_2(\nu_b)}{\Gamma_2(\omega_a) \Gamma_2(\nu_b)} \right) \left( 1 + \frac{2 i \chi_1}{\Gamma_1(\omega_a) + \Gamma_1(\omega_b)} \right)^{-1} \frac{\chi_1 \gamma_1^2}{\Gamma_1(\nu_b)\Gamma_1(\nu_a) \Gamma_1(\omega_b) \Gamma_1(\omega_a )} \notag \\
& \hphantom{\times \bigg[} + i \left( \frac{\conj{\Gamma}_1(\nu_a) \conj{\Gamma}_1(\omega_b)}{\Gamma_1(\nu_a) \Gamma_1(\omega_b)} \right) \left( 1 + \frac{2 i \chi_2}{\Gamma_2(\omega_a) + \Gamma_2(\omega_b)} \right)^{-1} \frac{\chi_2 \gamma_2^2}{\Gamma_2(\nu_b)\Gamma_2(\nu_a) \Gamma_2(\omega_b) \Gamma_2(\omega_a )} \bigg].
\end{align*}

\subsubsection{$N$ sites, counter-propagating photons}

For the $N$-site case, we only computed the S-matrices in the counter-propagating arrangement, and under the assumption of translation invariance, i.e.\ $\gamma_i = \gamma$, $\Delta_i = \Delta$, and $\chi_i = \chi$. From this, we obtain the single-photon S-matrix
\begin{equation*}
S_{\rm act,1}(\omega_k;\nu_k) = \left(- \frac{\conj{\Gamma}(\omega_k)}{\Gamma(\omega_k)} \right)^N \delta(\omega_k  - \nu_K),
\end{equation*}
for $k=a,b$, and the two-photon S-matrix is

\begin{align*}
S_{\rm act,2}(\omega_a, \omega_b;\nu_a, \nu_b)= & S_{\rm act,1}(\omega_a;\nu_a)S_{\rm act,1}(\omega_b;\nu_b) - i \frac{\chi \gamma^2}{\pi} \left(1 + \frac{2 i \chi}{\Gamma(\nu_b) + \Gamma(\nu_a)}\right)^{-1} \frac{\delta(\omega_a + \omega_b - \nu_a - \nu_b)}{\Gamma(\nu_b)\Gamma(\nu_a) \Gamma(\omega_b) \Gamma(\omega_a )} \notag \\
& \hphantom{\braket{\omega_a^{-}|\nu_a^{+}} \braket{\omega_b^{-}|\nu_b^{+}}} \times \bigg[ \sum_{j=1}^{N}
\left( \frac{\conj{\Gamma}(\omega_a)\conj{\Gamma}(\nu_b)}{\Gamma(\omega_a)\Gamma(\nu_b)} \right)^{N-j}
\left( \frac{\conj{\Gamma}(\omega_b)\conj{\Gamma}(\nu_a)}{\Gamma(\omega_b)\Gamma(\nu_a)} \right)^{j-1} \bigg].
\end{align*}


\section{Problem statement and rules for passive CPHASE gate via cross-Kerr interactions}\label{apx:problem_and_rules}

In this section we update Gea-Banacloche's~\cite{GeaBana10} suggested requirements for proposals of CPHASE gates. Paraphrasing, Gea-Banacloche's~\cite{GeaBana10} requirements were: (GB1) clearly local and physically realizable Hamiltonians; (GB2) localized wave packets using quantized multimode fields; (GB3) report gate fidelities computed; and (GB4) a realistic estimate of any residual losses or decoherence mechanisms.

We are particularly interested in determining if it is possible to construct a passive CPHASE gate using cross-Kerr interactions with the photons entering the device synchronously. By passive, we mean there should be no control pulses applied to the medium (atoms) unless they are static, and no active error correction.

We suggest the following requirements for ``in principle'' theory proposals
\begin{enumerate}
\item [P1:] Microscopic model. 

Ideally, the medium should be constructed from a microscopic model using clearly local, passive, and physically realizable Hamiltonians. If the model is a hybrid model, e.g. light coupling to a spin wave, one must actually model the coupling efficiency of this interface. This is comparable to (GB1).

One point which is often overlooked is that results obtained using particular phenomenological models cannot be extended as ``physical principles'', since a different medium could be constructed, out of different microscopic components, for which that result does not apply. 

\item [P2:] Multimode analysis.

The analysis should be done with a quantized multimode single photons, e.g.\ $\ket{1_\xi}= \int d\nu \xi(\nu) a\dg(\nu)\ket{0}$ where $[a(\omega),a\dg(\nu)]=\delta(\omega-\nu)$. If possible report the relevant S-matrices or the output wave functions. This is an extension to (GB2).

\item [P3:] Reflection must be explicitly dealt with. 

Either the proposal should use circulators or isolators to remove reflection altogether, or it should be shown theoretically that reflection can be suppressed enough to obtain a high-fidelity operation. For example, in proposals using continuous mediums, like atomic vapours, reflection is not usually modelled. 

\item [P4:] Quality of the operation.

Standard figures-of-merit are the average gate fidelity and the diamond norm between the proposal and the CPHASE gate, restricted to the encoded state space. This is comparable to (GB3).
 
\item [P5:] Imperfections. 

After making a convincing case for a particular theory proposal, one should then take imperfections into account. Examples include loss out of guided modes and coupling atoms (Kerr medium) to a thermal reservoir. However, it is important to keep in mind that the final word in the feasibility of any proposal is that of experiments. This means that the final proposal should model all (and only) realistic sources of noise for a particular implementation, and that results derived from this procedure should not be generalized to other experimental implementations. Finally, at this point it might be worthwhile to determine the fault tolerance threshold for the proposed imperfect gate. These points are made in (GB4).

\end{enumerate}

We now also point out a few things that, in our view, might \textbf{not} be necessary.

\begin{enumerate}
\item [P6:] Perfect operation.

It is not necessary that the operation is perfect, or that it can be made arbitrarily good. The threshold theorem of fault-tolerant quantum computing guarantees that
there exist some \textbf{constant} threshold above which error correction can be used to perform arbitrarily-long computations (this is why the average gate fidelity or the diamond norm should be used, see point P4 above). The exact value of the threshold is dependent on the noise model which should include, among other effects, the imperfections from point P5 above. 

\item [P7:] Single-photon distortion.

It is preferable to obtain a high-fidelity operation without deformation of single-photon wave packets. But, as we argued in the main text, that might not be strictly necessary since there are a number of ways to work around it.

\end{enumerate}

\end{document}